\documentstyle[aps,prl,twocolumn,epsfig,floats]{revtex}

\begin{document}
\draft
%\preprint{HEP/123-qed}
\twocolumn[\hsize\textwidth\columnwidth\hsize\csname %
@twocolumnfalse\endcsname
\title{Nothing moves a surface: vacancy mediated surface diffusion}
\author{R. van Gastel$^{*}$, E. Somfai$^{\dag}$, S.~B. van Albada$^{*}$,
    W. van Saarloos$^{\dag}$ and J.~W.~M. Frenken$^{*}$}
\address{$^{*}$Kamerlingh Onnes Laboratory, Universiteit Leiden, P.O. Box 9504,
    2300 RA Leiden, The Netherlands\\
    $^{\dag}$Instituut-Lorentz, Universiteit Leiden, P.O. Box 9506,
    2300 RA Leiden, The Netherlands}
\date{\today}
\maketitle
\begin{abstract}
We report scanning tunneling microscopy observations,
which imply that {\em all} atoms in a close-packed copper surface
move frequently, even at room temperature. Using a low density of embedded
indium `tracer' atoms, we visualize the
diffusive motion of surface atoms.
Surprisingly, the indium atoms seem to make concerted, long jumps. Responsible for
this motion is an ultra-low density of surface vacancies, diffusing rapidly
within the surface. This interpretation is supported by a detailed analysis of
the displacement distribution of the indium atoms, which reveals a shape
characteristic for the vacancy mediated diffusion mechanism that we propose.
\end{abstract}


\vspace{0.3in}
]
\narrowtext

Mobility of close-packed metal surfaces is usually thought to be restricted to
the direct vicinity of steps, where atoms most easily `come and go'. The
diffusion of adatoms along or between steps leads to characteristic step
fluctuations. Also adatom and vacancy islands are known to move via
rearrangements at their perimeter. Many STM-studies have been devoted to the
mobility of surfaces. Most of these have focussed on the motion of steps
\cite{wil99,lag97,kui93}, islands \cite{mor95} or adsorbates \cite{bes96}.
Recently it has been suggested that also surface vacancies can be responsible
for self-diffusion of islands \cite{han97}. Unfortunately, there are no
experimental techniques available with both the spatial and temporal
resolution necessary to follow the diffusion of naturally occurring vacancies
in a close-packed metal surface.

Indium which is deposited on Cu(001) has been found to modify the epitaxial
growth of copper on this surface. Its presence results in layer-by-layer
growth instead of rough three-dimensional growth \cite{veg95}. After
deposition the indium atoms proceed to steps on the copper surface
\cite{bre92,bre93}. At temperatures just below room temperature they are
incorporated in the outermost layer on substitutional terrace sites. In this
study we have used indium atoms that are embedded within the first layer of a
Cu(001) surface to monitor the diffusion of surface atoms \cite{gas00}. Our
observations lead us to conclude that surface vacancies are responsible for
the mobility of the indium and that this close-packed metal surface is far
from static, even at room temperature.

The experiments were performed with a variable temperature scanning
tunneling microscope (STM) \cite{hoo98} in
ultra-high vacuum (UHV). A Cu single crystal of 99.999 \% purity was
mechanically polished parallel to the (001)-plane \cite{koper}. Prior to
mounting the crystal in the UHV system we heated it in an Ar/H$_2$ atmosphere
to remove sulfur impurities. The sample surface was further cleaned in UHV by
several tens of cycles of sputtering with 600 eV Ar ions and annealing to 675
K. After approximately every fifth cycle the surface was exposed to a few
Langmuir of O$_2$ to remove carbon from the surface. STM images showed a
well-ordered surface with terrace widths up to 8000 \AA. Small quantities of
clean indium were deposited on the surface from a Knudsen cell.

\begin{figure}
  \centering
  \epsfig{file=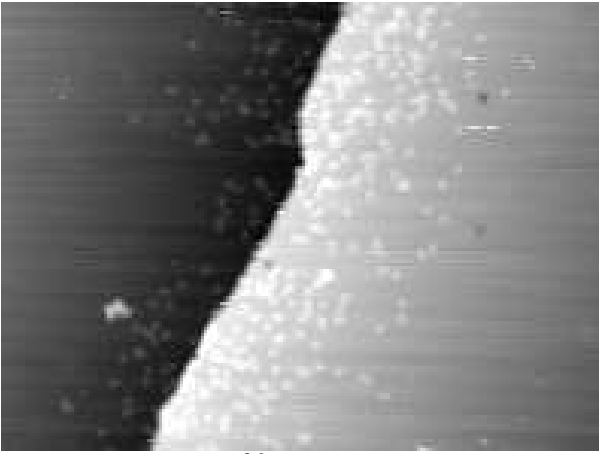,width=8.0cm}
  \hfill

  \caption
  {
    A 548 $\times$ 409 \AA$^2$ STM-image of a step on a Cu(001) surface, taken
42 minutes after deposition of 0.03 monolayer of indium at room temperature.
Embedded indium atoms show up as bright dots. The image shows a high density
of embedded indium atoms near the step. (I$_t$ = 0.1 nA, V$_t$ = -0.70 V)
  }
  \label{fig: stepimage}
\end{figure}

The starting point of the observations is shown in Fig.~\ref{fig: stepimage}.
At room temperature we have deposited 3\% of a monolayer of indium on the
Cu(001) surface. The STM image shows a region around an atomic step,
separating two flat terraces of the copper surface. The image was taken 42
minutes after deposition and shows that most indium atoms are within 150 {\AA}
from the step. From the apparent height of 0.4 {\AA} of the indium atoms, we
infer that they are embedded within the first copper layer. What we know from
lower-temperature STM experiments is that newly deposited indium atoms first
'hop' over the surface until they encounter a step. At the step they `invade'
the outermost copper layer (both on the upper and on the lower side of the
step), after which they diffuse away from the step, whilst remaining embedded
within the copper surface layer.

\begin{figure}[b]
  \centering
  \epsfig{file=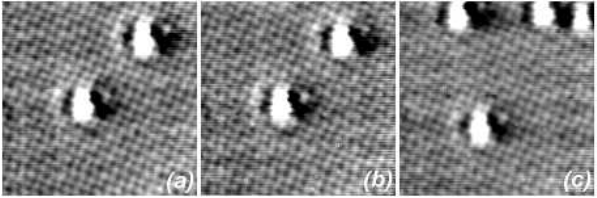,width=8.0cm}
  \hfill

  \caption
  {
    Three 50 $\times$ 50 \AA$^2$ STM-images selected from a movie measured at
RT illustrating the unusual diffusion of embedded indium atoms. In the time
interval of 160 s between images (a) and (b), no diffusion of the embedded
indium atoms has taken place. In image (c), taken 20 s later, a diffusion
event has taken place. Both indium atoms present in images (a) and (b)
have moved over several lattice spacings and two more indium atoms have
jumped into the imaged region. (I$_t$ = 0.9 nA, V$_t$ = -0.58 V)
  }
  \label{fig: stmmovie}
\end{figure}

\begin{figure}
  \centering
  \epsfig{file=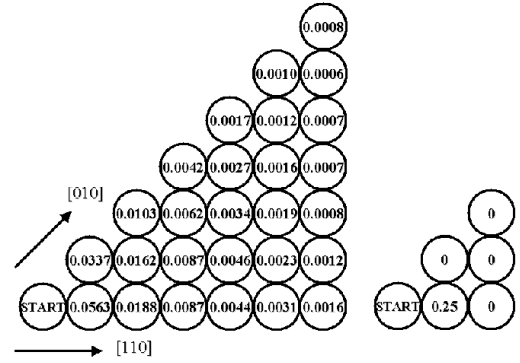,width=8.0cm}
  \hfill

  \caption
  {
    The distribution of jump vectors measured from STM-movies at 320
K. Plotted is the probability for jumps of an indium atom from its starting
position to each of the shown non-equivalent lattice sites. To illustrate the
unusual diffusion behavior, the expected jump vector distribution for the case
of simple hopping is plotted to the right.
  }
  \label{fig: jvd}
\end{figure}

We follow the diffusion of the embedded indium atoms in the copper terrace by
making series of images of the same area on the copper surface to form an
STM-movie of the motion \cite{gasweb}. To our initial surprise, we found that the
indium atoms move via long jumps of more than a single lattice spacing,
separated by long time intervals \cite{swa00}. In addition, the movies show
that there is a strong tendency for nearby indium atoms to jump at the same
time. Fig.~\ref{fig: stmmovie} illustrates this peculiar motion with a set
of three images taken from a movie measured at 320 K. From the STM movies we
have measured the distribution of jump lengths of the embedded indium atoms,
which has been plotted in Fig.~\ref{fig: jvd}. Note that there is a significant
probability for the indium atoms to make jumps as far as five lattice spacings.

The long jumps and the high probability of nearby indium atoms to jump
simultaneously, suggest strongly that diffusion of the indium is mediated by
another particle, which diffuses so rapidly that it remains invisible to the
STM.  The scenario that we propose is that the indium moves over several
lattice spacings during a multiple encounter with a single assisting particle
by changing places several times with that particle. The two obvious
candidates for this particle are adatoms (copper atoms on top of the surface
layer) and vacancies (missing atoms in the outermost copper layer). We can
rule out the first possibility on the basis of Fig.~\ref{fig: stepimage}. If
an indium atom were to change places with an adatom, it would thereby become
an adatom itself. We know from Fig.~\ref{fig: stepimage} and from other
observations that indium adatoms rapidly hop over the outermost copper layer
to the steps, without entering the copper surface directly. This means that if
an embedded indium atom would trade places with a copper adatom, it would
immediately disappear from the STM image and reappear somewhere at the step,
which is definitely not what we observe.

\begin{figure}[b]
  \centering
  \epsfig{file=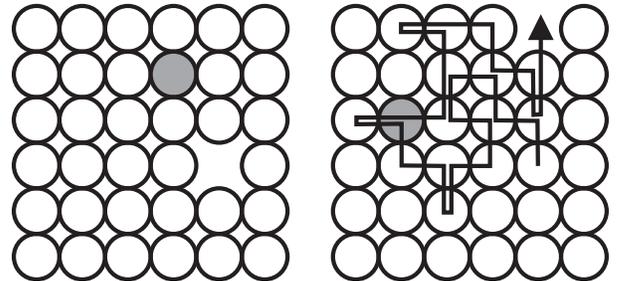,width=8.0cm}
  \hfill

  \caption
  {
    A ball model illustrating the vacancy-mediated diffusion mechanism
of the indium atoms.
  }
  \label{fig: cartoon}
\end{figure}

\begin{figure}
  \centering
  \epsfig{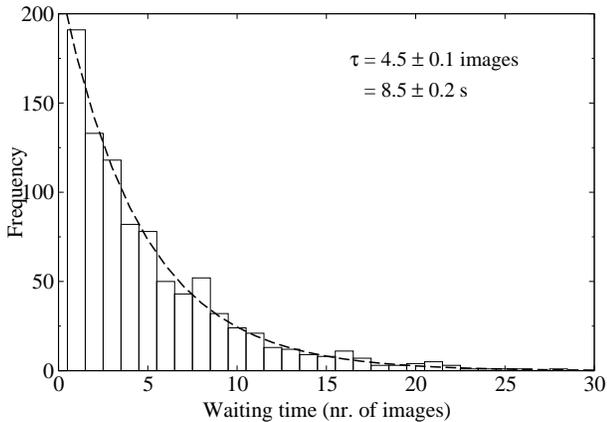}
  \hfill

  \caption
  {
    Time-interval statistics for subsequent jumps of individual indium atoms,
measured from STM-movies at 320 K with a time per image of 1.88 s. The dotted
curve is an exponential fit with a time constant $\tau = 8.5$ s.
  }
  \label{fig: wtd}
\end{figure}

Fig.~\ref{fig: cartoon} illustrates how a single surface vacancy can displace
an atom in the outermost copper layer, either an indium or a copper atom, over
several lattice spacings. In this mechanism, the length of the long jumps of
the indium atoms depends on the average number of times that a single vacancy
changes places with an indium atom, and we associate the frequency of the
(long) indium jumps with the frequency with which the indium is encountered by
new vacancies. We have measured the distribution of time-intervals between
consecutive jumps (see Fig.~\ref{fig: wtd}). The waiting time distribution is
purely exponential, from which we infer that individual long jumps are
uncorrelated in time and are therefore caused by different vacancies,
independently formed at random times. The fact that a single vacancy will
usually encounter various In atoms, naturally explains the tendency for nearby
indium atoms to jump at the same time.

The fact that we never see individual vacancies in the STM images and the fact
that the STM movies do not resolve the elementary steps in a
multi-lattice-spacing jump need not surprise us. Using the Embedded Atom Model
(EAM), we estimate that the formation energy of a vacancy in the Cu(001)
surface is 0.51 eV and that the activation energy for a surface atom to exchange
with the empty site, and thereby move the vacancy, amounts to 0.29 eV
\cite{longpaper}. Based on these estimates, we expect that at room temperature
only one surface atom out of roughly $6\cdot10^9$ is missing, and that each
empty site changes position with a high frequency, on the order of $10^8$ Hz.
These numbers are typical for close-packed metal surfaces and illustrate why it
is so difficult to see the vacancy diffusion at all. At low temperatures, where
vacancy motion would be slow enough to be followed by an inherently slow
instrument such as the STM, the probability of finding a vacancy is hopelessly
close to zero. At temperatures high enough for the surface to contain a
sufficiently high density of vacancies, the vacancies move much too fast to be
imaged at all.

In order to obtain a quantitative understanding of the jump vector
distribution of the embedded indium atoms, we performed a numerical calculation
as well as a continuum approximation, according to the following
model: The Cu(001) surface is a finite $l\times l$
square lattice, with copper atoms at the lattice sites; the boundary of
the lattice corresponds to the steps. One copper atom at the center of the lattice
is replaced by indium, and a vacancy is released one atomic site next to it.
The vacancy performs a biased random walk, its hopping probabilities to
the four different directions from each site are set from the diffusion
barriers calculated using the EAM potentials \cite{longpaper}.

The vacancy displaces the atoms in its path, including the indium atom.  When
the vacancy arrives at the boundary of the lattice, it is annihilated (it
recombines at the steps).  At this moment the displacement of the indium atom
is evaluated, and the whole process is repeated for the next vacancy to acquire
the distribution of the In jump vectors.

For the case of equal diffusion barriers and infinite lattices,
this problem has been solved analytically \cite{bru88}. Although the
results in some limits are quite similar to our continuum solution (see
below), the equal-barrier results are not directly applicable to the case
of indium in copper. Instead of moving isotropically, the vacancy neighboring
the indium atom has a much stronger preference to jump towards the indium
than to other directions, based on EAM barriers at room temperature (see below). This
difference has a significant impact on the indium jump distribution: the mean
square displacement is about 2.2 times larger than in the equal-barrier case, while
the overall shape of the distribution is about the same \cite{longpaper}.

It is computationally beneficial to separate the motion of the vacancy from
that of the indium atom.  For the indium atom, only the {\em direction} of the
next return of the vacancy is of importance, rather than the vacancy path which leads
to it. Therefore it is enough to calculate the probabilities of first return of
the vacancy to the indium atom from the four different directions after it left
the indium in one direction, as well as the probability of the vacancy's
recombination before return.  The In atom performs a random walk, where the
direction of each step with respect to the previous one is chosen according to
these return probabilities, and after each step the walk terminates with the vacancy's
recombination probability.  This procedure yields the proper final jump
distribution, while giving up the time information, which is experimentally
irrelevant anyway.  (This approach is valid under the assumption that the
environment of the indium does not change with the steps it takes, i.e., it is
still close to the middle of the lattice.)

\begin{figure}
  \centering
  \epsfig{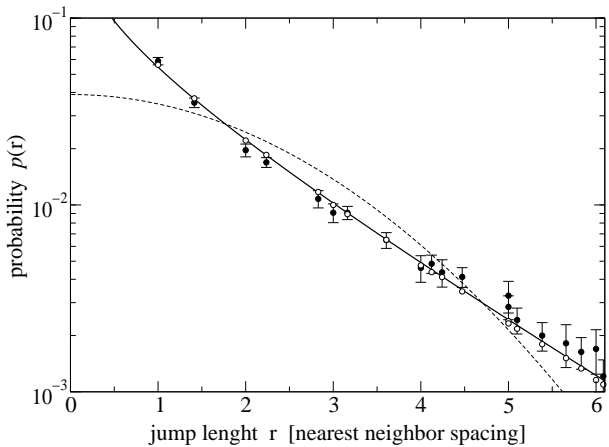}
\smallskip

  \caption
  {
    The distribution of the jump lengths of the indium atoms at 320 K. Filled
circles correspond to the experimental values, open circles are 
from the numerical calculation, and the solid curve is the isotropic continuum
(Bessel) approximation.
For comparison, the dashed curve shows the best-fit Gaussian distribution.
  }
  \label{fig: 1djvd}
\end{figure}

In practice, instead of using  Monte-Carlo type methods, we enumerate the possible
trajectories  to obtain the return probabilities and the indium jump vectors;
this provides superior convergence.  The following numerical values were calculated
for T = 320 K (EAM-barriers) and for a lattice size $l=401$, which corresponds to the typical
experimental terrace width of 1000 {\AA}. After leaving the indium atom to the right,
the vacancy's return probabilities from the four directions are the following:
$p_{\rm right}=1-2.4\cdot10^{-7}$, $p_{\rm up}=p_{\rm down}=1.1\cdot10^{-7}$,
$p_{\rm left}=4.2\cdot10^{-9}$, and the vacancy recombines with probability
$p_{\rm rec}=1.1\cdot10^{-8}$.  These values depend very weakly on the
lattice size $l$. The fact that two dimensions is the marginal dimension for
the return problem of a random walker implies a logarithmic $l$ dependence
of $p_{\rm rec}$. The root mean square jump length of the In atoms
is 3.5 nearest neighbor spacings.  The full distribution of the In jump
lengths is plotted in Fig.~\ref{fig: 1djvd} together with the experimental
values. The quantitative agreement supports our interpretation of the
mechanism of the indium diffusion.

Finally, we show that a simple continuum approach to this problem gives a
quite good approximation to the jump statistics. Let us denote the probability
of ``mobile
indium'' at position ${\bf r}$ with $\varrho({\bf r}, n)$, where the counter
$n$ measures the number of times the vacancy returns to the In atom.  The
indium is considered ``mobile'' while the vacancy is still around, and
``immobile'' after the vacancy has recombined. 
The effective diffusion equation for $\varrho({\bf r}, n)$ is
\begin{equation}
\frac{\partial\varrho({\bf r}, n)}{\partial n} = D \nabla^2\varrho - \epsilon
    \varrho\,.
\end{equation}
The first term corresponds to the
vacancy mediated diffusion of the mobile indium (isotropic, in this continuum approach),
and the second term to the
recombination of the vacancy, which makes the indium immobile.  The solution
in case of Dirac-delta initial conditions at the origin is 
\begin{equation}
\varrho({\bf r}, n) = \frac{1}{4\pi Dn}e^{-\frac{r^2}{4Dn}-\epsilon n} \,.
\end{equation}

We are interested in the final, ``immobile'' distribution of In:
\begin{equation}
p({\bf r}) = \int_0^\infty \epsilon\varrho({\bf r}, n) \,dn = 
    \frac{1}{2\pi} \frac{\epsilon}{D} \,\,
    {\rm K}_0\!\left(\frac{r}{\sqrt{D/\epsilon}} \right) \,,
\end{equation}
where ${\rm K}_0$ is the modified Bessel function of order 0. The parameters
can all be calculated: $\epsilon$ is the recombination probability $p_{\rm rec}$
of the vacancy, and the effective diffusion coefficient $D$ can be calculated
from the return probabilities, as will be discussed in detail
elsewhere \cite{longpaper}.

This analytical solution gives a good
approximation of the jump length distribution, as shown on
Fig.~\ref{fig: 1djvd}. We emphasize that neither the results of
the numerical calculation, nor the continuum solution contains any fitting
parameter: everything is calculated from the EAM barriers, the temperature and
the average terrace width.

In conclusion, the diffusive motion of the indium atoms can be explained by
the presence of a low density of extremely mobile vacancies in the first layer
of the surface. This interpretation is supported by the shape of the
distribution of measured jump lengths. The root mean square jump
length can be reproduced accurately in calculations if we take into account
the chemical difference between the indium and copper atoms. The theory
further shows that the multiple encounter of a single vacancy with a copper
atom in a clean copper surface should result in a root mean square displacement of
the atom of 1.6 nearest neighbor spacings.
Combining this number with the observed average jump rate of the
embedded indium atoms, we calculate a diffusion coefficient for copper atoms
in a Cu(001) surface of 0.42 \AA$^2\cdot$s$^{-1}$. We see that close-packed
terraces of metal surfaces, such as Cu(001), cannot be considered as static,
even at room temperature. The naturally occurring vacancies lead to a
continuous reshuffling of the surface, as if it were an atomic realization of
a slide puzzle!

We gratefully acknowledge B. Poelsema for help with the preparation of the
Cu-crystal. This work is part of the research program of the ``Stichting
voor Fundamenteel Onderzoek der Materie (FOM)'', which is financially
supported by the ``Nederlandse Organisatie voor Wetenschappelijk Onderzoek
(NWO)''.

\end{document}